# Using Heavy Clique Base Coarsening to Enhance Virtual Network Embedding

Ashraf A. Shahin[1,2]

[1]College of Computer and Information Sciences,
Al Imam Mohammad Ibn Saud Islamic University (IMSIU)
Riyadh, Kingdom of Saudi Arabia

[2]Department of Computer and Information Sciences, Institute of Statistical Studies & Research,
Cairo University,
Cairo, Egypt

*Abstract*—**Network virtualization allows cloud infrastructure providers to accommodate multiple virtual networks on a single physical network. However, mapping multiple virtual network resources to physical network components, called virtual network embedding (VNE), is known to be non-deterministic polynomial-time hard (*NP*-hard). Effective virtual network embedding increases the revenue by increasing the number of accepted virtual networks. In this paper, we propose virtual network embedding algorithm, which improves virtual network embedding by coarsening virtual networks. *Heavy Clique* matching technique is used to coarsen virtual networks. Then, the coarsened virtual networks are enhanced by using a refined *Kernighan-Lin* algorithm. The performance of the proposed algorithm is evaluated and compared with existing algorithms using extensive simulations, which show that the proposed algorithm improves virtual network embedding by increasing the acceptance ratio and the revenue.**

## I. INTRODUCTION

In cloud computing data centers, virtualization is employed to accommodate multiple virtual networks (VNs) on a single substrate network (SN), and multiple virtual servers on a single physical server [1]. Consolidating multiple virtual servers from the same virtual network to a single physical server coarsens virtual network down to a few physical servers. Coarsening VN reduces the cost of embedding by eliminating the cost of embedding virtual links between virtual nodes on the same substrate node. Although, effective VN coarsening can improve the utilization of SN's resources and increase the acceptance ratio of VNs and the revenue of infrastructure providers, most of current virtual network embedding algorithms do not take into account VN coarsening [2, 3, 4, 5, 6, 7].

In this paper, we propose virtual network embedding algorithm, which coarsens virtual networks using Heavy Clique matching technique. Then, the coarsened virtual networks are enhanced by using a refined Kernighan-Lin algorithm. The performance of the proposed algorithm is evaluated and compared with existing algorithms using extensive simulations, which show that the proposed algorithm improves virtual network embedding by increasing the acceptance ratio and the revenue.

The rest of this paper is organized as follows. Section 2 gives a short overview of related work. Section 3 presents the VN embedding model and problem formulation. Section 4 describes the proposed algorithm. Section 5 evaluates the proposed VN embedding algorithm. Finally, we conclude in section 6.

## II. RELATED WORK

In the last few years, many algorithms have been proposed for efficient VNE. VN embedding problem is *NP*-hard, and finding optimal solution can only be found for small problem instances [8]. Therefore, several heuristic algorithms have been proposed to find a good solution [5, 6, 7, 9]. Some algorithms have been proposed to find exact VNE solutions to be used as optimal bound for the heuristic based VNE solutions [4, 10].

Zhu and Ammar [11] proposed two VN embedding algorithms. In the first algorithm, allocated substrate resources are fixed throughout the VN lifetime. The performance of the first algorithm is improved by using heuristics and adaptive optimization. In the second algorithm, allocated substrate resources are reconfigured to increase the utilization of the underlying substrate resources. However, the proposed algorithms deal only with VNRs that are previously known and do not deal with VNRs that dynamically arrive over time.

In [12], Lischka and Karl proposed online VNE algorithm, which maps nodes and links during the same stage. The proposed algorithm maps VN to a sub-physical network that is similar to the topology of the VN and achieves previously defined constraints (e.g. CPU capacity, link bandwidth). During nodes mapping process, virtual nodes are sorted in descending order based on its required CPU and mapped sequentially to substrate nodes without allowing coexisting multiple virtual nodes from the same VN on one substrate node. To minimize the mapping cost, virtual links are mapped to substrate paths with minimal hops by incrementally increasing the maximum hop limit. However, the computational complexity of the proposed algorithm is high due to multiple operations. In [13], Di et al. improved performance and complexity of the proposed algorithm in [12] by considering the cost of mapping links during the process of sorting virtual nodes and choosing the maximal hop limit. Fischer et al. [3] modified the algorithm proposed in [12] to consider energy efficiency during nodes and links mapping. Fischer et al. allowed mapping several virtual nodes of the same virtual network to the same substrate node. Although, they take into account the energy efficiency during consolidating virtual nodes, they did not consider the mapping cost.

In [10], Cheng et al. proposed two-stage VN embedding algorithm, called RW-MaxMatch, which ranks nodes using topology-aware node ranking technique to reflect the topological structure of the VNs and the SN. However, RW-MaxMatch algorithm maps nodes without considering its relation to the link mapping, which leads to high consumption of the underlying SN's resources. This is due to mapping neighboring virtual nodes widely separated in the SN.

In [10], Cheng et al. improved the coordination between nodes and links mapping in the RW-MaxMatch algorithm by proposing RW-BFS algorithm. RW-BFS algorithm is a backtracking one-stage VN embedding algorithm, which maps nodes and links at the same stage. In [14, 15], Zhang et al. proposed two VN embedding models: an integer linear programming model and a mixed integer-programming model. To solve these models, Zhang et al. proposed an enhanced version of the MaxMatch algorithm, called RW–PSO algorithm, based on particle swarm optimization. RW–PSO algorithm reduces the time complexity of the link mapping stage by using shortest path algorithm and greedy k-shortest paths algorithm.

To improve the coordination between nodes mapping stage and links mapping stage, Chowdhury et al. [16, 17] formulated the VNE problem as a mixed integer program (MIP), which is NP-hard. To obtain polynomial-time solvable algorithms, they relaxed the integer program to linear program, and proposed two VNE algorithms: D-ViNE (deterministic VNE algorithm) and R-ViNE (randomized VNE algorithm). Nogueira et al. [18] proposed heuristic-based VN embedding algorithm to deal with the heterogeneity of VNs and SN, in both links and nodes. The proposed algorithm is one stage VNE algorithm.

Some of existing works proposed VN embedding algorithms to embed VNRs in distributed cloud computing environments [19, 20, 21, 22]. Houidi et al. [23] proposed exact and heuristics VN embedding algorithms, which split virtual network requests using max-flow min-cut algorithms and linear programming techniques. Leivadeas et al. [24] proposed VN embedding algorithm based on linear programming.

The proposed algorithm partitions VNRs using partitioning approach based on Iterated Local Search. Houidi et al. [25] proposed distributed VN embedding algorithm, which is performed by agent-based substrate nodes. The authors proposed VN embedding protocol to allow communication between the agent-based substrate nodes. However, the proposed algorithm deals only with the offline VN embedding problem.

## III. VIRTUAL NETWORK EMBEDDING MODEL AND PROBLEM FORMULATION

*Substrate network (SN):* We model the substrate network as a weighted undirected graph $G_s = (N_s, L_s)$, where $N_s$ is the set of substrate nodes and $L_s$ is the set of substrate links. Each substrate node $n_s \in N_s$ is weighted by the CPU capacity, and each substrate link $l_s \in L_s$ is weighted by the bandwidth capacity. Fig. 1(b) shows a simple SN example, where the available CPU resources are represented by numbers in rectangles and the available bandwidths are represented by numbers over the links.

*Virtual network (VN):* virtual network $VN_i$ is modeled as a weighted undirected graph $G_{v_i} = (N_{v_i}, L_{v_i})$, where $N_{v_i}$ is the set of virtual nodes and $L_{v_i}$ is the set of virtual links. Virtual nodes and virtual links are weighted by the required CPU and bandwidth, respectively. Fig. 1(a) shows an example of VN with required CPU and bandwidth.

*Virtual network requests (VNR):* the $i^{th}$ VN request $vnr_i$ in the set of all VN requests $VNR$ is modeled as $(G_{v_i}, t_{a_i}, t_{l_i})$, where $G_{v_i}$ is the required VN to be embedded, $t_{a_i}$ is the arrival time, and $t_{l_i}$ is the lifetime. When $vnr_i$ arrives, substrate nodes' CPU and substrate links' bandwidth are allocated to achieve the $vnr_i$. If the substrate network does not have enough resources to achieve $vnr_i$, $vnr_i$ is rejected. At the end of $vnr_i$ lifetime, all allocated resources to $vnr_i$ are released.

*Virtual Network Embedding (VNE):* embedding $VN_i$ on SN is defined as a map $M: G_{v_i} \rightarrow (N_s', P_s')$, where $N_s' \subseteq N_s$, and $P_s' \subseteq P_s$, where $P_s$ is the set of all loop free substrate paths in $G_s$. Embedding $VN_i$ can be decomposed into node and link mapping as follows:

Node mapping: $M_N: N_{v_i} \rightarrow N_s'$

Link mapping: $M_L: L_{v_i} \rightarrow P_s'$

For example, mapping of the VN in Fig. 1(a) on SN in Fig. 1(b) can be decomposed into:

Node mapping: $\{a \rightarrow B, b \rightarrow A, c \rightarrow C\}$

Link mapping: $\{(a, b) \rightarrow \{(B, A)\}, (b, c) \rightarrow \{(A, D), (D, C)\}, (c, a) \rightarrow \{(C, B)\}\}$

*Virtual Network Embedding Revenue:* as in [8, 10, 14], the revenue of embedding $vnr_i$ at time $t$ is defined as the sum of all required substrate CPU and substrate bandwidth by $vnr_i$ at time $t$.

$$R(vnr_i, t) = Life(vnr_i, t). \left( \sum_{n_{v_i} \in N_{v_i}} CPU(n_{v_i}) + \sum_{l_{v_i} \in L_{v_i}} BW(l_{v_i}) \right)$$

Where $CPU(n_{v_i})$ is the required CPU for the virtual node $n_{v_i}$, $BW(l_{v_i})$ is the required bandwidth for the virtual link $l_{v_i}$, and $Life(vnr_i, t) = 1$ if $vnr_i$ is in its lifetime and substrate resources are allocated to it, otherwise $Life(G_{v_i}, t) = 0$.

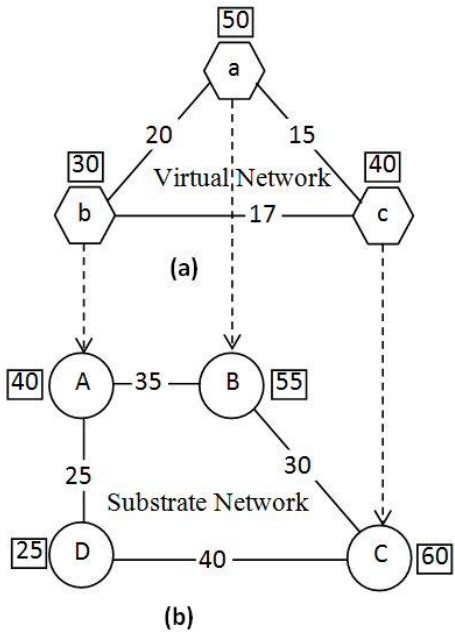

Fig. 1. Example 1 of VNE

*Virtual Network Embedding Cost*: as in [8, 10, 14], the cost of embedding $vnr_i$ at time $t$ is defined as the sum of all allocated substrate CPU and substrate bandwidth to $vnr_i$ at time $t$.

$$Cost(vnr_i, t) = Life(vnr_i, t).\left(\sum_{n_{v_i} \in N_{v_i}} CPU(n_{v_i}) + \sum_{l_{v_i} \in L_{v_i}} BW(l_{v_i}).Length(M_{L_{v_i}}(l_{v_i}))\right) \quad (1)$$

Where $Length(M_{L_{v_i}}(l_{v_i}))$ is the length of the substrate path that the virtual link $l_{v_i}$ is mapped to.

*Objectives*: the main objectives are to increase the revenue and decrease the cost of embedding virtual networks in the long run. To evaluate the achievement of these objectives, we use the following metrics:

- *The long-term average revenue*, which is defined by

$$\lim_{T \to \infty} \left(\frac{\sum_{t=0}^{T} \sum_{i=1}^{I} R(vnr_i, t)}{T}\right) \quad (2)$$

Where $I = \parallel VNR \parallel$, and $T$ is the total time.

- *The VNR acceptance ratio*, which is defined by

$$\frac{\parallel VNR_s \parallel}{\parallel VNR \parallel} \quad (3)$$

Where $VNR_s$ is the set of all accepted virtual network requests.

- *The long term R/Cost ratio*, which is defined by

$$\lim_{T \to \infty} \left(\frac{\sum_{t=0}^{T} \sum_{i=1}^{I} R(vnr_i, t)}{\sum_{t=0}^{T} \sum_{i=1}^{I} Cost(vnr_i, t)}\right) \quad (4)$$

## IV. THE PROPOSED ALGORITHM

In this section, we describe the motivation behind the proposed algorithm and describe the details of the proposed algorithm, which is called HCM-VNE algorithm.

### A. Motivation

VN embedding cost (defined by equation 1) depends on allocated substrate CPU and allocated substrate bandwidth. VN embedding cost can be reduced by minimizing these resources. However, minimizing allocated substrate CPU may violate service level agreement and reduce the quality of the service provided to the customers. Allocated substrate bandwidth can be reduced by increasing the number of virtual links between virtual nodes that are mapped to the same substrate node. VN embedding cost is reduced by eliminating the cost of embedding such virtual links. However, finding VN embedding solution with maximum number of eliminated virtual links is not easy task. For example, to map VN in Fig. 2(a) to SN in Fig. 2(b), Fig. 2 shows the mapping solution with the maximum number of eliminated virtual links among other solutions. This solution can be reached by finding sub-VNs that are close to be clique and map each sub-VN to one substrate node. This example motivates us to propose HCM-VNE algorithm, which coarsens VNs using heavy clique matching technique before mapping it.

### B. The HCM-VNE algorithm

Algorithm 1 shows the steps of the proposed HCM-VNE algorithm. In line 1, $CPU_{max}$, which is the upper bound of the coarsened node CPU, is set to the maximum available CPU in SN. In line 2, the upper bound of the total coarsened node bandwidth, $BW_{max}$, is set to the maximum available bandwidth in SN. VNs are coarsened using $Coarsening()$ function and coarsened VNs are optimized using $Optimize()$ function. $Coarsening()$ function and $Optimize()$ function will be described later on. The HCM-VNE algorithm constructs breadth-first searching tree for the graph of the coarsened VN. The root node of the constructed tree is the coarsened virtual node with the largest resources (sum of CPU and BW). Nodes in each level in the created breadth-first searching tree are sorted in descending order based on their resources. Finally, in line 8, the HCM-VNE algorithm embeds coarsened VN on SN using $Embed()$ function.

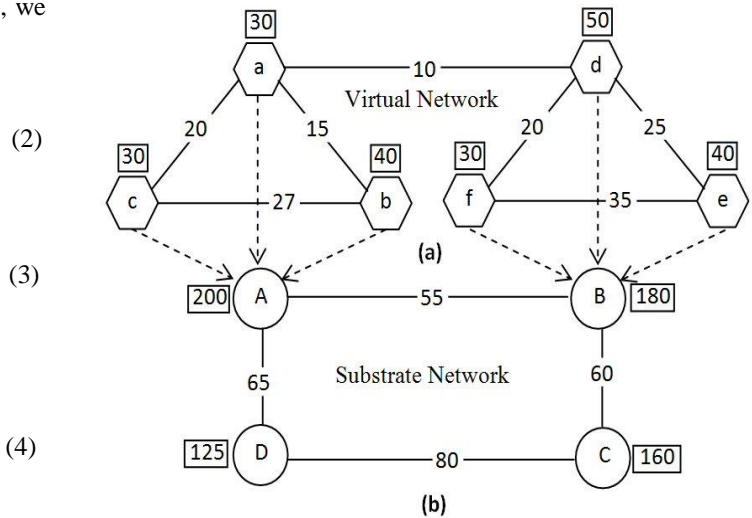

Fig. 2. Example 2 of VNE

| ALGORITHM 1: The details of the HCM-VNE algorithm |
| --- |

**INPUTS**:

$G_v = (N_v, L_v)$: VN to be embed

$G_s = (N_s, L_s)$: SN to embed on

$Max\_hops$: maximum allowed substrate path length

$Max\_backtrack$: upper bound of nodes re-mapping operation

**OUTPUTS**:

$M(G_v)$: map VN nodes and links to SN's resources

$S\_VNE$: VN embedding success flag

**Begin**

1: $CPU_{max} = Max_{n_s \in N_s}(CPU(n_s))$

2: $BW_{max} = Max_{n_s \in N_s}(\sum_{l_s \in L'_s} BW(l_s))$,

   where $L'_s \subseteq L_s$ and $n_s$ is incident in each $l_s \in L'_s$

3: $G_c = Coarsening(G_v, CPU_{max}, BW_{max})$

4: $G_c = Optimize(G_c, CPU_{max}, BW_{max})$

5: Build breadth-first searching tree of $G_c$ from coarsened virtual node with largest resources.

6: Sort all nodes in each level in the created breadth-first tree in descending order according to their required resources.

7: backtrack_count=0

8: **if** $Embed(G_{c_{root}}, G_s, M(G_v))$ then

9:     $S\_VNE = true$

10:    **return**

11: **else**

12:    $S\_VNE = false$

13:    **return**

14: **end if**

**End**

## C. Coarsening() function

Virtual networks are coarsened using heavy clique matching technique. A clique in undirected graph is a fully connected subgraph. The cost of embedding VNs is reduced by embedding each sub-VN that is close to clique on one substrate node.

To determine how close sub-VN $G'_v = (N'_v, L'_v)$ is to a clique, we define link density $L_{density}(G'_v)$ as

$$L_{density}(G'_v) = 2\|L'_v\|/(\|N'_v\|(\|N'_v\| - 1))$$

If the sub-VN $G'_v$ is clique (or fully connected), the number of edges is equal to $(\|N'_v\|(\|N'_v\| - 1))/2$ and the link density $L_{density}(G'_v)$ goes to one. $L_{density}(G'_v)$ is small if the sub-VN $G'_v$ is far from being clique.

Algorithm 2 shows the details of the *Coarsening()* function. Coarsening process is iterative and starts with an initial coarsening graph $G_c = (N_c, L_c)$, which is created and initialized by creating coarsened node for each virtual node and coarsened link for each virtual link. Each coarsened node $n_{c_i} \in N_c$ can be considered as a sub-VN $G_{v_i} = (N_{v_i}, L_{v_i})$, where $N_{v_i} \subseteq N_v$ (at this time each $N_{v_i}$ contains only one virtual node), and $L_{v_i} \subseteq L_v$, such that each virtual link $l_{v_i} \in L_{v_i}$ connects two virtual nodes in $N_{v_i}$. Each coarsened link $l_{c_i} \in L_c$ between two coarsened nodes is a set of virtual links connect virtual nodes in these coarsened nodes. Each virtual node exists in exactly one coarsened node, and each virtual link exists in exactly one coarsened node or one coarsened link. For example, VN in Fig. 2(a) can be coarsened as in Fig. 3.

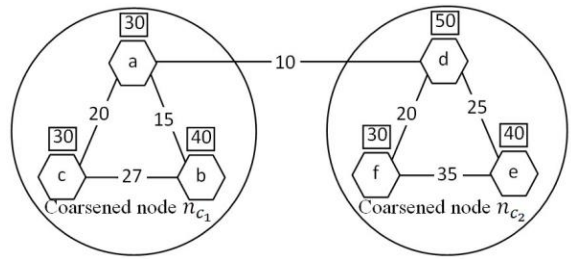

Fig. 3.    Coarser VN for the VN in Fig. 2(a)

The graph of the coarsened VN in Fig. 3 is

$G_c = (\{n_{c_1}, n_{c_2}\}, \{l_{v_1}\})$, where

$n_{c_1} = (\{a, b, c\}, \{(a, b), (b, c), (c, a)\})$,

$n_{c_2} = (\{d, e, f\}, \{(d, e), (e, f), (f, d)\})$, and

$l_{v_1} = \{(a, d)\}$

In *Coarsening()* function, coarsened nodes are visited in a sequential way, and each unmatched coarsened node $n_{c_i}$ is matched with its unmatched neighbor $n_{c_j}$ such that the new coarsened node created by combining $n_{c_i}$ and $n_{c_j}$ achieves the CPU and BW constraints and its $L_{density}$ is the largest among all possible coarsened nodes created by combining $n_{c_i}$ with other unmatched neighbors. If such neighbor exists, we add coarsened node $n_{c_i}$ with its neighbor $n_{c_j}$ to the matching list $M_{List} = \{(n_{c_i}, n_{c_j}) \mid n_{c_i}, n_{c_j} \in N_c \text{ and } n_{c_i}, n_{c_j} \text{ are matched}\}$. At the end of each iteration, coarser graph $G_c$ is updated by combining each pair in $M_{List}$ to a new coarsened node. If $M_{List}$ is empty the *Coarsening()* function terminates.

## D. Optimize() function

*Coarsening()* function coarsens VN in Fig. 2(a) as in Fig. 3. However, *Coarsening()* function combines coarsened nodes only based on link density and does not consider the required bandwidth for each virtual link, which sometimes increases the cost of VN embedding. For example, if the virtual link $(a, d)$ in Fig. 3 has bandwidth equal to 50, coarser VN can be improved by moving the virtual node $a$ from the coarsened node $n_{c_1}$ to the coarsened node $n_{c_2}$. Fig. 4 shows the optimized coarsened VN.

To optimize coarsened VN, we used a refined Kernighan-Lin (KL) algorithm. In 1970, Kernighan-Lin (KL) algorithm was proposed by Kernighan and Lin for graph partitioning problem. Kernighan-Lin (KL) algorithm partitions graph into two parts with equal sizes and with minimal number of cutting edges. It starts with an initial bipartition of the graph and searches for two subsets of vertices from each part of the graph, such that they have the same number of vertices and swapping them improves the cost of the partition. Kernighan-Lin algorithm swaps the selected subsets and repeats the entire process until no such subsets found [26]. However, standard Kernighan-Lin algorithm deals only with typical graph partitioning problem, so it is not directly applicable to optimize coarsened VNs, which may be partitioned to more than two partitions with different sizes.

ALGORITHM 2: The details of the *Coarsening*() function

**INPUTS:**
  $G_v$: VN graph to be coarsened
  $CPU_{max}$: the upper bound of the coarsened node CPU
  $BW_{max}$: the upper bound of the total coarsened node BW
**OUTPUTS:**
  $G_c$: coarsened VN graph
**Begin**
1: Create and initialize coarsening graph $G_c = (N_c, L_c)$
2: Create new matching list $M_{List}$
3: **while**(true)
4:    **for** each unmatched coarsened node $n_{c_i} \in N_c$
5:       Find unmatched neighbor $n_{c_j} \in N_c$ such that
          $CPU\left(n_{c_i} \cup n_{c_j}\right) \leq CPU_{max}$,
          $BW\left(n_{c_i} \cup n_{c_j}\right) \leq BW_{max}$, and
          $L_{density}\left(n_{c_i} \cup n_{c_j}\right) = Max_{n_{c_k} \in N'_c}\left(L_{density}\left(n_{c_i} \cup n_{c_k}\right)\right)$,
          *where* $N'_c$ *is the set of all neighbors of the node* $n_{c_i}$
6:       Add $\left(n_{c_i}, n_{c_j}\right)$ to $M_{List}$
7:    **end for**
8:    **if** $M_{List} == \emptyset$ **then**
9:       **break**
10:   **else**
11:      Update $G_c$ by combining each pair in $M_{List}$
12:         $M_{List} = \emptyset$
13:   **end if**
14: **end while**
**End**

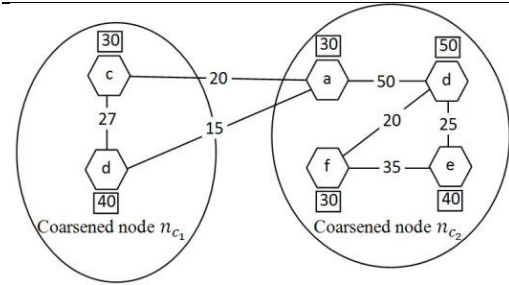

Fig. 4.   Optimized coarser VN for the coarser VN in Fig. 3

ALGORITHM 3: The details of the *Optimize*() function

**INPUTS:**
  $G_c = (N_c, L_c)$: coarsened VN to be optimized
  $CPU_{max}$: upper bound of the coarsened node CPU
  $BW_{max}$: upper bound of the Total coarsened node BW
**OUTPUTS:**
  $G_c$: optimized coarsened VN
**Begin**
1: Terminate=false
2: **while** (NOT Terminate)
3:    Terminate=true
4:    **for** each $n_{c_i} \in N_c$
5:       **for** each boundary virtual node $n_v \in n_{c_i}$
6:          **if** $\exists\ n_{c_j} \in N_c$, such that $\sum_{l_v \in l_v^*} BW(l_v) > \sum_{l_v \in l_v^*} BW(l_v)$,
             where
             $l_v^*$ is the set of all virtual links between $n_v$ and virtual nodes in $n_{c_j}$,
             and
             $l_v^*$ is the set of all virtual links between $n_v$ and virtual nodes in $n_{c_i}$
7:          **then**
8:             **if** moving $n_v$ from $n_{c_i}$ to $n_{c_j}$ does not violate CPU and BW
                constraints.
9:             **then**
10:               Move $n_v$ from $n_{c_i}$ to $n_{c_j}$
11:               Terminate=false
12:            **else**
13:               Find set of boundary virtual nodes $n'_v$ in the coarsened node
                  $n_{c_j}$, such that swapping $n_v$ and $n'_v$ improves bandwidth and
                  does not violate CPU and BW constraints.
14:               **if** such node found swap them
15:                  Terminate=false
16:               **end if**
17:            **end if**
18:         **end if**
19:      **end for**
20:   **end for**
21: **end while**
**End**

### E. Embed() function

The *Embed()* function embeds coarsened VN on SN as described in algorithm 4. In the *Embed()* function, candidate substrate node list for each coarsened virtual node is built by collecting all substrate nodes that have available CPU capacity at least as large as the coarsened virtual node CPU and have a loop free substrate path to each substrate node contains one of the previously mapped neighbors. Each substrate path should satisfy the constraint of the maximum substrate path length, and have available bandwidth greater than or equal the bandwidth of the coarsened virtual link between the coarsened virtual node and its previously mapped neighbor.

Candidate substrate nodes for each coarsened virtual node are collected by creating a breadth-first search tree from each substrate node contains one of the previously mapped neighbors, and finding the common substrate nodes between the created trees. In the constructed trees, substrate nodes should satisfy the CPU constraints for coarsened virtual node, and substrate paths should satisfy the connectivity constraints to connect the coarsened virtual node with its neighbors. By this way, all candidate substrate nodes in the candidate

To optimize coarsened VN, we redefined Kernighan-Lin (KL) algorithm as shown in algorithm 3. *Optimize()* function starts with the partition performed by the *Coarsening()* function and moves boundary virtual nodes between coarsened nodes to improve edge-cut, such that this movement does not violate the CPU and BW constraints. Virtual node is called boundary node, if it is connected to virtual nodes outside its coarsened node. For example, in Fig. 3, virtual node $a$ is a boundary virtual node for the coarsened node $n_{c_1}$, because it has virtual link to the virtual node $d$, which is not in the coarsened node $n_{c_1}$.

If moving the selected boundary virtual node to the target coarsened node violates the CPU or BW constraints, we try to find one or more boundary nodes in the target coarsened node to be swapped with the selected boundary virtual node. If no such boundary virtual node is found, we postpone this movement and recheck it again in the next iteration. The whole process is repeated until no movements are performed.

substrate node list satisfy all constraints (CPU and connectivity constraints).

Substrate nodes in the candidate substrate node list are sorted in ascending order according to the total cost of embedding coarsened virtual links from the coarsened virtual node to all previously embedded neighbors. If the coarsened virtual node is a root node, the candidate substrate node list is a set of all substrate nodes that have enough resources to embed the coarsened virtual node. The candidate substrate nodes for the root are sorted in descending order according to the total available resources.

Coarsened virtual node is sequentially mapped to substrate nodes in its candidate substrate node list. If there is no appropriate substrate node in its candidate substrate node list, we backtrack to the previously mapped node, re-map it to the next candidate substrate node, and continue to the next node. In line 3, mappings of the coarsened virtual node and its coarsened virtual links are added to $M(G_v)$ by using the function $Add()$. To map coarsened node $n_{c_i}$ to substrate node $n_s$, the function $Add()$ adds maps from each virtual node in $n_{c_i}$ to the substrate node $n_s$. All virtual links in the coarsened node $n_{c_i}$ are mapped to substrate paths with length zero from the substrate node $n_s$ to itself. For each coarsened link from $n_{c_i}$ to one of the previously mapped coarsened nodes, the function $Add()$ adds maps for all virtual links in these coarsened links. Virtual links are mapped to shortest loop free substrate paths, which are specified by breadth-first search manner. In line 6, $Delete()$ function is used to perform the backtracking process.

---

**ALGORITHM 4:** The details of $Embed()$ Function

**INPUTS:**

    $n_{c_i}$: current coarsened virtual node to be embedded

    $G_s$: substrate network to embed on

    $M(G_v)$: map of the previously mapped nodes and links

**OUTPUTS:**

    $M(G_v)$: updated map

    $S\_VNE$: VN embedding success flag

**Begin**

1: Build candidate substrate node list $C_i$ for $n_{c_i}$

2: **for** each $n_s$ in $C_i$

3:    Add$\left((n_{c_i}, n_s), M(G_v)\right)$

4:    **if** Embed($n_{c_{i+1}}$, $G_{s_i}$, $M(G_v)$) **then return true**

5:    **else**

6:        Delete$\left((n_{c_i}, n_s), M(G_v)\right)$

7:    **end if**

8:    **if** backtrack_count > Max_backtrack **then return false**

9: **end for**

10: backtrack_count ++

11: **return false**

**End**

---

## V. PERFORMANCE EVALUATION

We evaluated the proposed HCM-VNE algorithm by comparing its performance with some of existing algorithms.

First, we implemented three algorithms: HCM-VNE, RW-MaxMatch [15], and RW-BFS [10]. Second, we generated SN topology and 3000 VN topologies to be used as inputs to the implemented algorithms. Finally, we compared the results from the implemented algorithms. In the following sub-sections, we describe the evaluation environment settings and discuss the results of the simulations.

### A. Evaluation environment settings

In our evaluation, the substrate network topology is configured to have 200 nodes with 1000 links. Substrate network is generated using Waxman generator. Bandwidths of the substrate links are real numbers uniformly distributed between 50 and 100 with average 75. We have selected two server configurations: HP ProLiant ML110 G4 (Intel Xeon 3040, 2 cores X 1860 MHz, 4 GB), and HP ProLiant ML110 G5 (Intel Xeon 3075, 2 cores X 2660 MHz, 4 GB). Each substrate node is randomly assigned one of these server configurations.

Virtual network topologies are generated using Waxman generator with average connectivity 50%. Number of virtual nodes in each VN is variant from 2 to 20. Each virtual node is randomly assigned one of the following CPU: 2500 MIPS, 2000 MIPS, 1000 MIPS, and 500 MIPS, which are correspond to the CPU of Amazon EC2 instance types. Bandwidths of the virtual links are real numbers uniformly distributed between 1 and 50. VN's arrival times are generated randomly with arrival rate 10 VNs per 100 time units. The lifetimes of the VNRs are generated randomly between 300 and 700 time units with average 500 time units. 3000 VN topologies are generated and stored in brite format. For each algorithm, we run the simulation for 30000 time units with the previously generated VNRs[1]. For all algorithms, we set the maximum allowed hops (Max_hops) to 2, and the upper bound of remapping process (Max_backtrack) to $3n$, where $n$ is the number of nodes in each VNR.

### B. Evaluation results

Three metrics have been used to evaluate the performance of the proposed algorithms: *the long-term average revenue*, which is defined by Equation (2), *the VNR acceptance ratio*, which is defined by Equation (3), and *the long-term R/Cost ratio*, which is defined by Equation (4). Fig. 5 shows the simulation results using the VNR acceptance ratio to compare the different VNE algorithms. It can be seen that the proposed algorithm that coarsened VNs using heavy clique matching increases the acceptance ratio compared with other algorithms. For example, at time unit 30000, in Fig. 5, the VNR acceptance ratio for the RW-BFS and RW-MaxMatch are 20 and 16 percent, while the VNR acceptance ratio for the HCM-VNE is 53 percent. In other words, the proposed algorithm can embed more VNs on the same SN at the same time. Consequently, the proposed algorithm increases the long-term average revenue compared with other algorithms, as shown in figure 6.

---

[1]The generated SN topology, generated VNRs topologies, and outputs are available online at (https://drive.google.com/folderview?id=0BxEBmTQ 0WG5RcnBYLVZhdW42bjg&usp=drive_web)

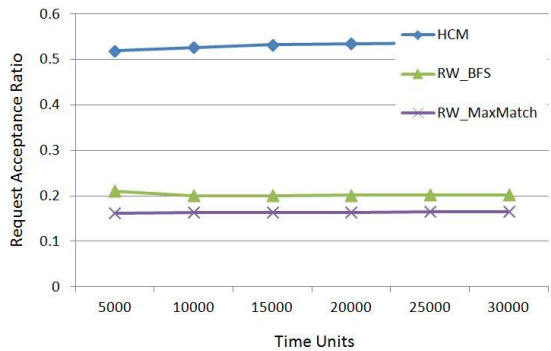

Fig. 5. The VNR acceptance ratio comparison

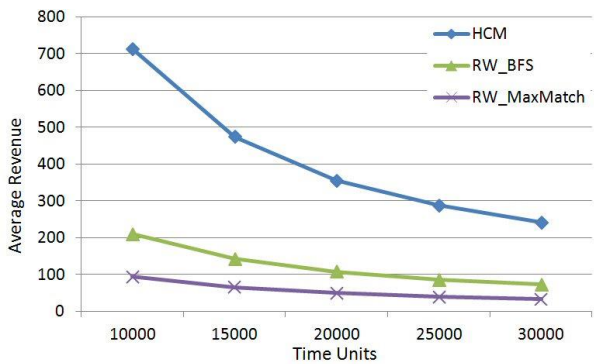

Fig. 6. The long-term average revenue comparison

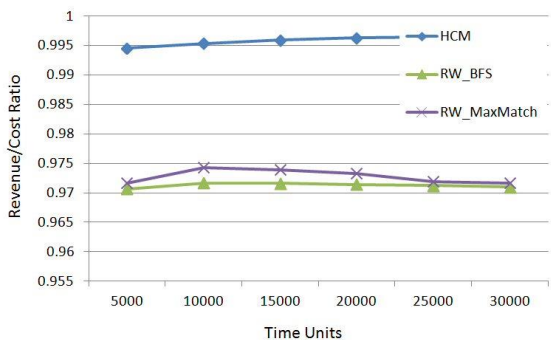

Fig. 7. The long-term Revenue/Cost ratio comparison

For example, at time unit 30000, the average revenue for the RW-BFS and RW-MaxMatch are 72 and 33, while the average revenue for the HCM-VNE is 240. As shown in Fig. 7, the *long-term Revenue/Cost ratio* of all algorithms are nearly the same, but the proposed algorithm performs slightly better than other algorithms.

## VI. CONCLUTION

In this paper, we proposed virtual network embedding algorithm, which coarsens virtual networks using heavy clique matching and optimizes the coarser virtual networks by applying a refined Kernighan-Lin (KL) algorithm. The proposed algorithm coarsens sub-virtual networks that are close to clique and embeds each sub-virtual network to substrate node. The cost of embedding virtual networks is reduced by eliminating the cost of embedding virtual links between virtual nodes on the same substrate node. Performance of the proposed algorithm has been evaluated and compared with some of the existing algorithms using extensive simulations. Extensive simulation experiments show that the proposed algorithm increases the acceptance ratio and the revenue. For the future work, we plan to investigate other coarsening techniques (e.g. Random Matching and Light Edge Matching) to find the best coursing technique, which increases the acceptance ratio and the revenue while decreasing the embedding cost.